\documentclass[doublecol]{epl2} 
\usepackage{amsmath}

\usepackage{graphicx}
\usepackage{dcolumn}
\usepackage{bm}

\title{The Derivation of the Fundamental Superconducting Interaction and the
Phase Diagram of Cuprates}
\shorttitle{The Derivation of the Superconducting Interaction of Cuprates}

\author{E. V. L. de Mello}

\institute{
  \inst{1} Instituto de F\'isica, Universidade Federal Fluminense, Niter\'oi, RJ 24210-340, Brazil\\
}

\abstract{
We show that an electronic phase separation (EPS) transition described by the
Cahn-Hilliard theory yields regions of low free energy forming grains of
low and high charge densities. These local differences in the potential energy 
are studied numerically and used here for the first time
as the origin of an attractive interaction that gives rise to
local pair formation. The resistivity transition $T_c$
occurs due to Josephson coupling among these superconducting
regions or grains. Using this approach 
within the Bogoliubov-deGennes calculations we derive the
Gaussian shape of $T_c$ against the entire doping level $p$.
We find that every pairing gap develops locally at a 
temperature $T_p$, following the relation $2\Delta/k_BT_p\approx8.0$
in close agreement recent measurements.}

\begin{document}

\maketitle
\section{Introduction}

There are increasing evidences that the charge
distribution in the $CuO_2$ planes of the high temperature superconductors
(HTSC) is microscopically inhomogeneous. Several different experiments like
neutron diffraction\cite{Tranquada,Bianconi,Bozin}, muon spin relaxation ($\mu SR$)\cite{Uemura,Sonier},
NQR and  NMR\cite{Singer,Curro} have  detected a position dependent
electronic density.
These experiments indicated that such disorder is
stronger on the underdoped side of the phase diagram
and it is possibly related
with the non Fermi liquid behavior of the normal phase.
However, recent STM studies on Bi2212 reveal
spatial variations of the  electronic gap amplitude on a nanometer length scale even
on overdoped compounds\cite{McElroy,Gomes,Pasupathy}. These data on
overdoped samples, which behave as normal
Fermi liquids at high temperatures, indicates that a phase separation
transition may occur at temperatures not much larger than $T_c$. A transition
at such low temperatures (of the order of 100K) is likely 
to be an intrinsic electronic
phase separation and not due to ionic mobility.

On the theoretical side there are many predictions 
that hole carriers may segregate into hole-rich and 
hole-poor regions at low doping due to strong carrier-carrier
correlation effect\cite{Gorgov,Yukalov,EK}. 

Here we deal with a
completely different approach, namely, a phase separation transition driven
by the minimization of the free energy\cite{Mello09}. As the temperature goes below the
pseudogap temperature\cite{TS,Tallon}
 $T^*$, the phase separation (PS) process starts.
Such second order phase transition is
a direct explanation to the NQR measurements\cite{Singer} of
two different signals coming from two different types of
local doping domains as the
temperature goes down. The origin of this EPS transition is the
proximity to the insulator AF phase, common to all
cuprates, as we derived from the principle of the competing
minimum free energy\cite{Mello09}. When the temperature decreases
below $T_{PS}(p)$, the free energy of the
homogeneous system with an average doping level or 
charge density $p$ becomes {\it higher} than the 
disordered one, made mainly of two values of the local charge concentration at each
point $\vec r_i$.  According the stripe phase measurements\cite{Tranquada,Bianconi} 
and the NQR data\cite{Singer}, 
the local values of the density $p(\vec r_i)\equiv p(i)$ follows a bimodal 
distribution~\cite{Mello03} formed of AF domains
with $p(i) \approx 0$ and high hole density domains with
$p(i)\approx 2p$. These domains, clusters or grains in the Cu-O planes are 
of nanometer size containing 10-100 sites.
\section{The Electronic Phase Separation}
To trace the EPS and the cluster formation we use the
general theory of Cahn-Hilliard (CH)\cite{CH}. It
describes how a system evolves from small fluctuations around the
average charge concentration $p$ near the phase separation temperature
$T_{PS}(p)$ to a complete separation into low and high
density grains, passing by intermediates configurations
as the temperature decreases.
The order parameter of such transition is the difference between the
temperature dependent local charge or doping concentration
$p(i,T)$ and the average doping level $p$, i.e., $u(i,T)\equiv (p(i,T)-p)/p$.
The Ginzburg-Landau (GL) free energy functional
in terms of $u(i,T)$ for a given compound $p$
near the transition temperature is given by

\begin{eqnarray}
f(i,T)= {{{1\over2}\varepsilon^2 |\nabla u(i,T)|^2 +V(u(i,T))}}\equiv K+V_{GL}.
\label{FE}
\end{eqnarray}
Where the potential ${\it V}_{GL}(p,i,T)= A^2(T)u^2/2+B^2u^4/4+...$,
$A^2(T)=\alpha(T_{PS}(p)-T)$, $\alpha$ and $B$ are temperature
independent parameters. $\varepsilon$ gives
the size of the grain boundaries among two distinct
phases\cite{Otton,Mello04}.
$V_{GL}(p,i,T)$ changes with
the site position "i"; $V(u=0)=0$ at the grain boundaries where $u=0$, and has two minima
$V_{GL}(u,t)= -A^4(T)/B$ in the lowest and highest density
sites where the order parameter assumes the
value $u_{min}(T)=\pm \sqrt{A(T)/B}$. These two  ${\it V}_{GL}$minima 
regions work as attractors for the holes, giving rise to 
the clusters and to the main point here; it creates 
an effective two-body attractive potential.

The CH equation can be written\cite{Bray} in the
form of a continuity equation of the local free energy $f(i,T)$,
$\partial_tu=-{\bf \nabla.J}$, with the current ${\bf J}=M{\bf
\nabla}(\delta f/ \delta u)$, where $M$ is the mobility or the
charge transport coefficient. Therefore,
\begin{eqnarray}
\frac{\partial u}{\partial t} = -M\nabla^2(\varepsilon^2\nabla^2u
+ A^2(T)u-B^2u^3).
\label{CH}
\end{eqnarray}
We have already made a detailed study of the  CH differential equation by
finite difference methods\cite{Otton} which yields the density profile
$u(p,i,T)$ in a $105\times 105$ array as function of the time steps,
up to the stabilization of the local densities, using parameters in the CH
simulation that
yield stripe\cite{DDias07,DDias08} and patchwork\cite{Mello04,Mello08,Caixa07}
patterns at intermediate time regimes.

\section{The Fundamental Interaction}
Here we introduce a new approach in order to derive the attractive
potential that segregates the holes into grains of low and high
density and which will be used as an effective two-body attraction 
to calculate the superconducting properties. The justification to 
this procedure is because, given the mesoscopic size of a grain and 
the low doping values of the HTSC, there are very few holes in each
confined region.

\begin{figure}[!ht]
\includegraphics[height=7cm]{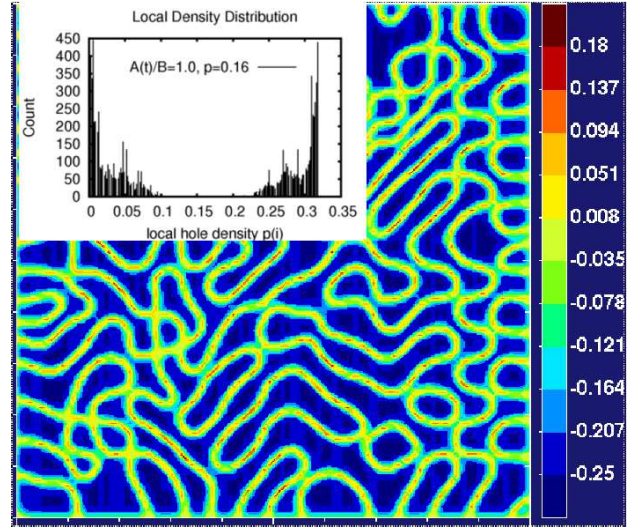}
\caption{(color online) The $V_{GL}$ potential energy map showing the
regions of minimum potential separated by the lines of grain boundaries. The inset is
the local charge density histogram for $p=0.16$ showing the tendency toward
a bimodal charge distribution. The points marked below
the inset are where we made the local density of states 
calculations presented in the end. }
\label{FE6100}
\end{figure}

Consequently we follow numerically the
"kinetic" and potential energy ($V_{GL}$) map, that are the first and second
term in Eq.(\ref{FE}). These energy maps show
that the grains, either the low and high density ones, are regions
of free energy minimum, as shown in Fig.(\ref{FE6100}).
This figure describes a common intermediate regime  of disorder, between
a homogeneous system with a Gaussian distribution of densities and
a bimodal distribution for a complete phase separation, as
it is shown in the inset by the histogram of local densities $p(i)$.

Following the $V_{GL}$ values, from low temperatures
up to $T_{PS}(p)$ when the grains disappear completely, we can obtain
the qualitative behavior of this attractive potential. Here we
use parameters to reproduced the Bi2212 low temperature data
of McElroy et al\cite{McElroy}
\begin{eqnarray}
V_{GL}(p,i,T)=V(p) V(T)=(-0.9+2.8p)|u(i,T)|^2,
\label{VpT}
\end{eqnarray}
where the values are in $eV$, $V(p)$ is linear and vanishes
at $p\approx 0.32$ following the behavior $T^*(p)$ or
$T_{PS}(p)$. $V(T)$ vanishes at $T_{PS}(p)$ and
increases as T decreases. The $u(i,T)$ is also
parameterized in a similar way, $|u(i,T)|^2=u(i)(1-T/T_{PS})^{(3-T/T_{PS})}$ 
showing how the system becomes homogeneous at $T_{PS}$, and the
other limit at $T=0$ of largest phase separation
with $A=B=1$.

The energy minimum at the grains and the grain boundary energy
barrier keep the holes confined forming metallic clusters with 
typical Fermi energies much smaller than the bulk Debye energy, the so called
anti-adiabatic regime\cite{JR97}. This condition with the effective
hole attraction toward the cluster center is highly
favorable to bipolaron formation\cite{JR97}. Then
the effective hole attraction, modelled by $V_{GL}$, is taken
as {\it the origin of the superconducting
interaction} in the form of a local two-body attraction in the
Bogoliubov-deGennes (BdG) theory. This assumption is in
agreement with the observation that a large electron
mobility and a small Fermi energy produces a large Nernst
Effect\cite{Behnia} as observed in many cuprates\cite{Ong}.

\section{BdG-CH Combined Calculations}
To calculate the local superconductivity amplitude or gap 
function $\Delta_d(i,T)$ in a plane, we use the BdG theory with
the extended Hubbard Hamiltonian in a similar fashion
as we did before for a phenomenological next neighbor
potential\cite{Mello04,DDias07,Mello08,Caixa07,DDias08}.
Except from the temperature dependent $V_{GL}(p,i,T)$, introduced here
for the first time, and the granular charge density taken from
the CH calculations and used as an input,
all the others parameters are similar to
values previously used in other calculations\cite{Mello04,DDias07,DDias08,Caixa07,Mello09}.

The {\it BdG-CH combined calculations} yield large superconducting
amplitudes in high density grains and
low or almost zero local amplitudes at regions with low densities.
We define the {\it local superconducting temperature} $T_c(i)$
as the onset temperature for $\Delta_d(i,T)$. 
The largest value of $T_c(i)$ in a given compound
determines the temperature
$T_{on}(p)$ which marks the onset of superconductivity. Since
$T_{on}(p)$ is correlated with the potential $V(p,i,T)$
it is larger in the underdoped region and decreases almost linearly
as the average doping increases, similar to
the Nernst effect onset temperature\cite{Ong,DDias07} or a 
signal above $T_c$ associated also with pair formation by
tunneling measurements\cite{Moura}.

As the temperature decreases below $T_{on}(p)$  some
grains become superconductors, the zero resistivity transition takes
place when the Josephson coupling $E_J$ among these grains is
sufficiently large to overcome thermal fluctuations\cite{Merchant}. Thus
$E_J(p,T=T_c) \approx k_BT_c(p)$ what leads to phase locking and
long range phase coherence. Consequently the  superconducting
transition in cuprates occurs in two steps, similar to a
superconducting material embedded  in a non superconducting
matrix\cite{Merchant}: First by the appearing of intragrain
superconductivity and by Josephson coupling with phase locking at a
lower temperature, {\it what provides a clear interpretation to the
presence of the
superconducting amplitude $\Delta_d(i,T)$ above $T_c(p)$}. 
\section{The Cuprates Phase Diagram}

In this approach $T_c(p)$ is not directly related with the
intragrain superconductivity, and the 
amplitudes $\Delta_d(i,T)$ do
not change appreciably around $T_c(p)$, specially for underdoped
compounds, characterized by very large $T_{on}(p)$. 
This consequence of the present granular
theory has been experimentally verified by temperature dependent
tunneling\cite{Suzuki} and angle resolved photon
emission\cite{Campuzano}.
\begin{figure}[ht]
\begin{center}
  \begin{minipage}[b]{.1\textwidth}
    \begin{center}
     \centerline{\includegraphics[width=8.0cm]{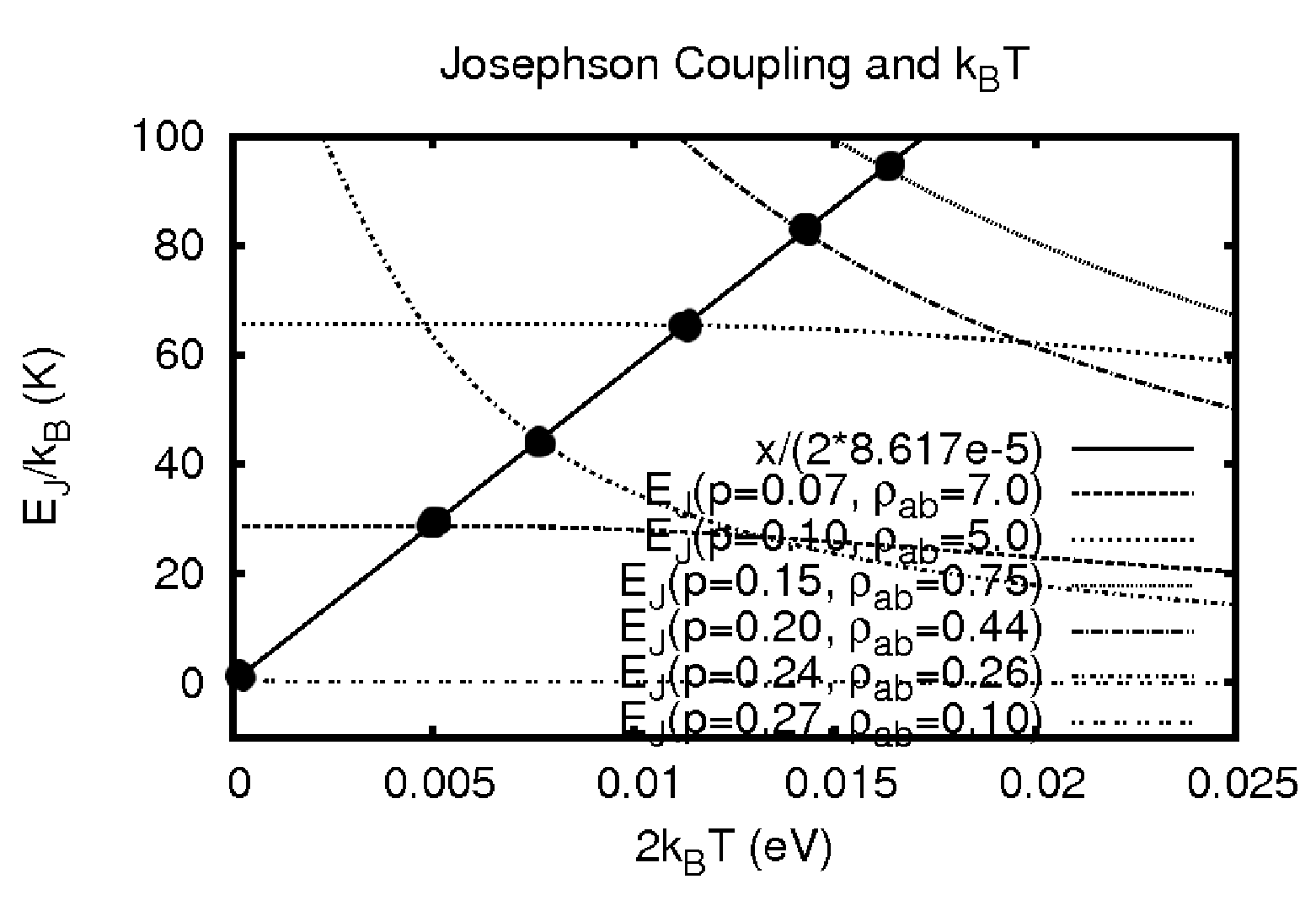}}
    \centerline{\includegraphics[width=8.0cm]{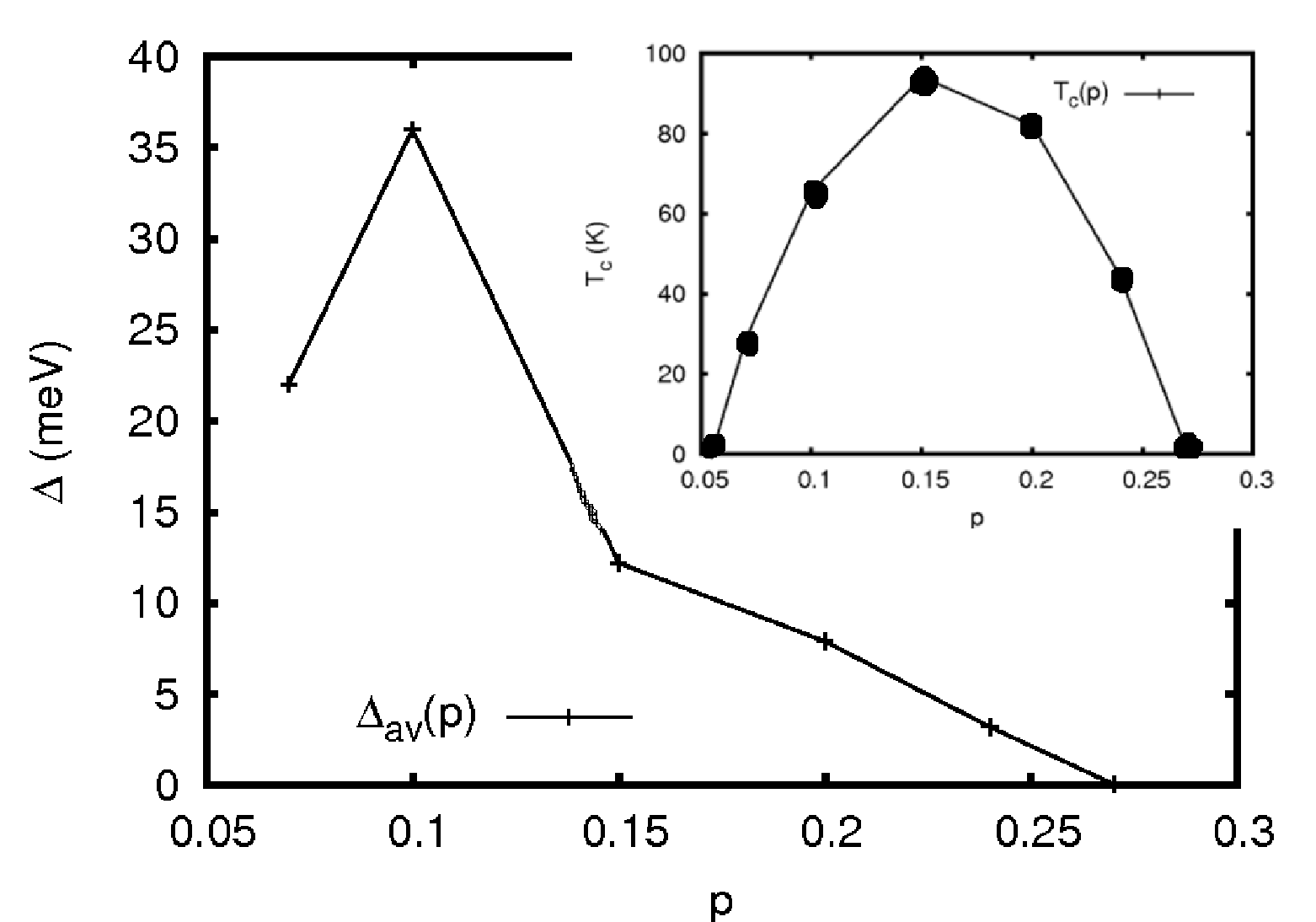}}
    \end{center}
  \end{minipage}
\caption{In the top panel we plot
$k_BT$ and the Josephson coupling among superconducting grains $E_J(p,T)$
for some selected doping values as function of T. The curves intersections
give the dome shaped $T_c(p)$, as plotted in the inset in the
down panel where we show the average $\Delta(T,p)$ 
calculations.  }
\label{EJTc}
 \end{center}
\end{figure}
Using to the theory of granular superconductors\cite{AB} to
the electronic grains,
\begin{eqnarray}
E_J(p,T) = \frac{\pi h}{4 e^2 R_n} tanh(\frac{\Delta(T,p)}{2K_BT_c}).
\label{EJ}
\end{eqnarray}
Where $\Delta(T,p)$ is the average of the BdG superconducting gaps
calculations $\Delta_d(i,T)$ on a $24 \times 24$ square taken from 
the $105 \times 105$ mesh after the CH simulations as 
that shown in Fig.(\ref{FE6100}).
The $R_n$ is the normal resistance of a
given compound, which we take as proportional to the $\rho_{ab}$
measurements\cite{Takagi} on the complete series  of
$La_{2-p}Sr_pCuO_2$. These $R_n$ values are given in the legend of the 
Fig.(\ref{EJTc}) top
panel. The average $\Delta(T,p)$ as function
of $p$ are plotted in the low panel of Fig.(\ref{EJTc}) with
the $T_c(p)$ results from Eq.{\ref{EJ}) are in
the inset and yields the well known dome shape with
excellent agreement with the Bi2212 values.
This is one of the most important result of our CH-BdG calculations.
\begin{figure}[ht]
\begin{center}
  \begin{minipage}[b]{.1\textwidth}
    \begin{center}
     \centerline{\includegraphics[width=5.5cm,angle=-90]{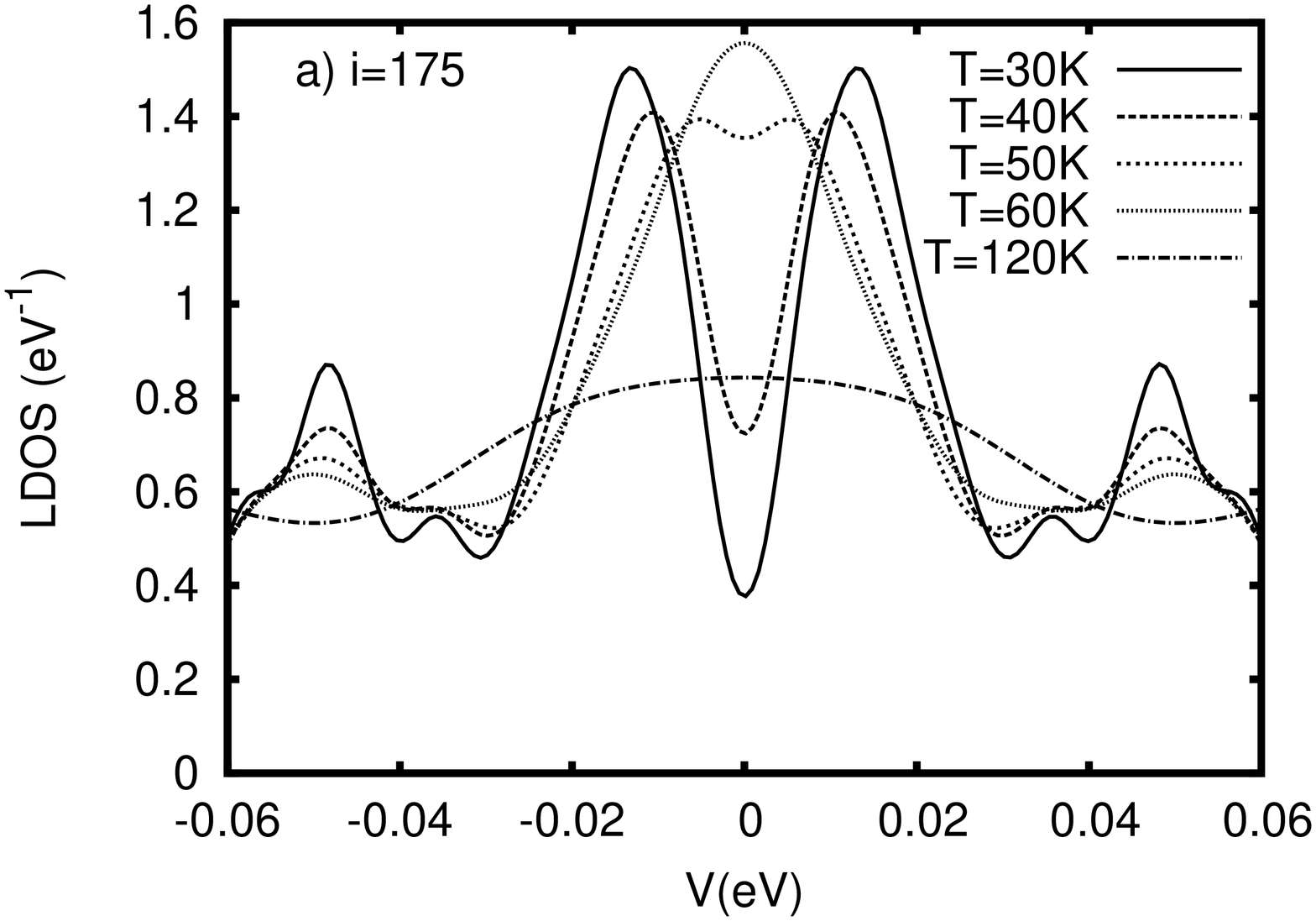}}
    \centerline{\includegraphics[width=8.0cm]{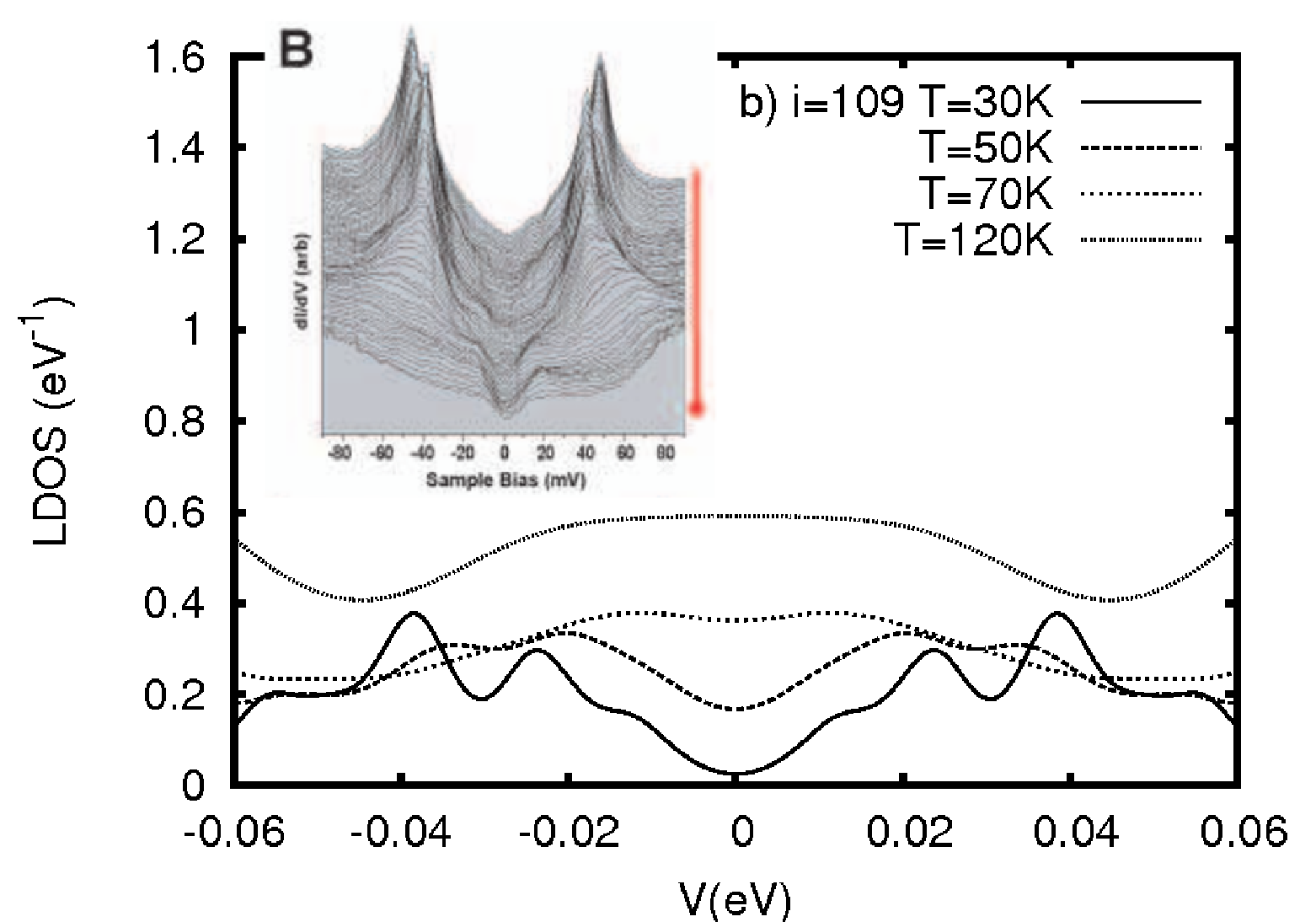}}
    \end{center}
  \end{minipage}
\caption{In the panel a) we plot the LDOS at a "metallic" site (i=175) 
inside a high density grain with $\Delta(T=30K)=17meV$. As the temperature
increases the coherent gaps close (at $T\approx 52$K). 
In panel b) we show a LDOS with at an "insulator" 
grain (i=109) which has both $p(i)$ and $\Delta (T=0)$ almost
zero. At $T=120$K both LDOS converges to
the same value all over the plane because the proximity 
to $T_{PS}=140$K. The inset shows the low temperature STM data of
McElroy et al\cite{McElroy} displaying the same behavior of our calculations. }
\label{DETp}
\end{center}
\end{figure}
\section{The STM Results and Interpretation} 

According to the recent STM data\cite{Gomes,Pasupathy},
the low values of $T^*(p) \sim T_{PS}(p)$ for overdoped samples rules out any
ionic mobility as the origin of the inhomogeneities in overdoped samples
and probably at all dopings.
Thus we want to show that these STM results can be interpreted by
the granular behavior resulting from the EPS, calculating
the symmetric local density of states (LDOS) at different 
local doping,
\begin{eqnarray}
 N_i(E)=\sum_n[&|&u_n({\bf
x}_i)|^2 + |v_n({\bf x}_i)|^2]\times  \nonumber \\
&& [f_n^{'}(E-E_n)+f_n^{'}(E+E_n)],
\label{LDOS}
\end{eqnarray}
where $f_n$ is the Fermi function,
the prime is the derivative with respect to the argument, and $u_n, v_n$ and $E_n$
are respectively the eigenvectors and positive eigenvalues (quasi-particles 
exciting energy) of the BdG matrix
equation\cite{Mello04,DDias07,Caixa07,DDias08,Mello09}.

Here we concentrate on a $p=0.23$ compound
that is close to the Bi2212 compounds used in
recently STM experiments with many novel results\cite{Gomes,Pasupathy}.
The low temperature CH EPS calculations on this sample with 
the estimated value of $T_{PS}=140$K
yield that  $\approx 20\%$ of
the sites are in very low densities regions with $p(i)<0.05$. These low
density unitary cells  have quite distinct
LDOS behavior than the high density ones as it is shown in
Fig.(\ref{DETp}). 

The calculations at sites in the high density
regions have larger superconducting
amplitudes $\Delta (i,T)$ with well defined (coherent) 
and high LDOS peaks. $\Delta (i,T)$ is calculated directly from the BdG
equations but can be also estimated by the position of the first
LDOS peak. As the temperature
increases, it shows the closing of these peaks and the building up of
spectral weight at the Fermi level (E=0) (Fig.(\ref{DETp}a)). 
In particular, for this case, we
have a $\Delta(0,i=175)\approx 17.7$meV and the gap derived from the peaks
decreases and closes near $T=52$K. This behavior is quite common to
the sites with $p(i)>0.05$ and with $T_c(i)$
around 45K and 60K.

On the other hand, the calculations at points with $p(i)<0.05$
display a rather different behavior as it is shown in panel b). 
As an example we show the LDOS at $i=109$ with $p(i)\approx 0$ and 
$\Delta(0,i=109)\approx 0$. It shows a much smaller LDOS (compare
the vertical scale of panel a) and b)) without
the low temperatures coherent peaks
as in panel a) and with small oscillations far from the Fermi
energy given an impression of a larger (incoherent) gap.
These small oscillations followed by other peaks are quite
similar to the firsts few peaks in the work of McElroy et
al\cite{McElroy} showing in the inset of panel b). As the
temperature increases and reaches $T=50K$ we see that
the LDOS at panel a) is almost close yielding  a $\Delta(i=173,t=50K)\sim 0$
while the LDOS at panel b) has still a large dip that
gives a  $\Delta(i=109,t=50K)\sim 20$meV. This dip in the
LDOS stays much above the resistivity transition $T_c=60$K in
close agreement with the STM maps of Gomes et al\cite{Gomes}.

Since the $V_{GL}$ potential (Eq.(\ref{VpT})) vanishes with the 
order parameter, the LDOS for both cases converge to the same
value as the temperature approaches the "melting" temperature $T_{PS}=140$K,
as it is demonstrated in the Fig.(\ref{DETp}) by the $T=120$K curve.

Another striking result of the CH-BdG calculations is that, despite the uncertainty
on $T_c(i)$ (or $T_p$ in the notation of Gomes et al\cite{Gomes})
for very small gaps, mostly of our results
follow close the measured relation
$2\Delta_d/K_BT_c(i) \approx 7.9$\cite{Gomes}. For the case of
i=175, $T_c(i)=52K=4.5$meV and $2\Delta_d(i,0)=34$mev which
gives a ratio of 7.6. For i=109, taking the first peak in
Fig.(\ref{DETp}b) at 26meV which closes completely at $T\approx 74$K,
we obtain again $2\Delta /K_BT_c(i) \approx 8.1$ again in close
agreement with the STM data\cite{Gomes} but larger than the value
measured by tunneling of $2\Delta_d/K_BT_c(i) \approx 6.0$\cite{Moura}.
Since the
lower density (insulator) grains have much smaller LDOS around the
Fermi level, they also have
lower local conductivity, although the gaps measured by the position
of the LDOS peaks, as discussed above, are
larger than high density sites with their coherent peaks. 
Consequently regions with larger gaps have lower conductivity
as it was recently measured\cite{Pasupathy}.

\section{Conclusion}
In summary we have made a detailed study of the EPS in HTSC using
the CH theory that allows us to followed the local free energy minima
that generate
the high and low density grains in the Cu-O planes. 
The differences in these free energy local minima were calculated
numerically and used in the BdG approach as the superconducting
interaction in the electronic grains of cuprates in the non adiabatic limit. 
These calculations give rise to the intragrain 
superconductivity and they provide a scenario to the pseudogap phase,
as composed of local regions with finite superconducting amplitude
$\Delta(i,p,T)$ without phase locking.
The Josephson coupling calculations in connection with the measured values of
the resistivity yield an accurate $T_c(p)$ curve for the whole doping
values of the Bi2212 series.
The LDOS calculated in the high density grains yield the coherent
peaks, and those in the low density regions give larger gaps and
ill defined peaks in agreement with the STM data in Bi2212. As far
as we know, this present work is the only one to give an interpretation
to all the whole STM results.

All of these calculations in close agreement with current 
data led us to conclude that the
EPS is an important ingredient above the superconducting phase
and generates mostly of all the intricate normal phase physics of the
HTSC.

I gratefully acknowledge partial financial aid from Brazilian
agency CNPq.

\end{document}